\documentclass[twoside]{article} 
\usepackage{fleqn,espcrc2} 
\usepackage{psfig}  
 
\title{On the search for a narrow penta-quark Z$^+$ baryon in NN  
interactions\thanks{Supported 
by Forschungszentrum J\"ulich and the Australian Research  
Council}} 
\author{M.V. Polyakov$^{1,2}$, A. Sibirtsev$^3$, K. Tsushima$^4$,  
W. Cassing$^3$ and K. Goeke$^1$ \\ \vspace{3mm}  
$^1$Institut f\"ur Theoretische Physik II, Ruhr-Universit\"at  
Bochum, D-44780 Bochum, Germany \\  
$^2$ Petersburg Nuclear Physics Institute, Gatchina, 188350 Russia\\  
$^3$Institut f\"ur Theoretische Physik, Universit\"at Giessen,  
D-35392 Giessen, Germany \\  
$^4$Special Research Center for the Subatomic Structure of  
Matter (CSSM) and \\  
Department of Physics  and Mathematical Physics,  
University of Adelaide, SA 5005, Australia}   
\begin{document}    
\begin{abstract}  
The possibility for an observation of  
a narrow penta-quark $Z^+$  
baryon in $NN$ reactions is discussed. It is shown that the  
$pp{\to}n\Sigma^+K^+$ reaction at excess energies  
around 100~MeV above threshold provides optimal conditions  
for $Z^+$ baryon detection by an  analysis of the $nK^+$  
invariant mass spectrum, if the $Z^+$ mass is located around 1.5 GeV 
involving a rather narrow width.  
  
\vspace{1pc}  
\end{abstract}  
\maketitle   
  
The standard valence quark model builds up the low-lying 
baryons from three  valence quarks $qqq$ with strangeness from 
$S{=}{-}3$ to $S{=}0$, that are 
surrounded by a meson cloud or strong $q\bar{q}$ vacuum polarizations. 
However, also Fock states involving additional $q\bar{q}$ components are 
allowed which then can appear as baryonic resonances. The inverse 
life time of these excited states is proportional to the phase 
space for the decays allowed 
by quantum numbers. Interesting excitations are possible for a    
$qqqq\bar{q}$ configuration~\cite{MIT} which allows to  
construct $S{=}{+}1$ states denoted by $Z$ baryons.  
The observation of $Z$ baryons thus provides an unambiguous  
signal that the standard valence quark model  
has to be extended to a larger Fock space. 
  
An experimental search for $Z$ baryons was  
started  in 1966 at the BNL~\cite{Cool} by the observation of a  
{\it clear} resonance peak in the $K^+p$ and $K^+d$ total  
cross section at kaon momenta around 0.9$\div$1.3~GeV/c.  
This novel baryonic resonance with strangeness $S{=}{+}1$,  
mass $M_Z \approx$ 1.91~GeV and width $\Gamma_Z \approx$ 180~MeV 
was interpreted as  
a SU(3) antidecuplet together with the $N_{11}^\ast$(1480).  
  
A review on the further experimental and theoretical  
activities and an evidence for strangeness $S{=}{+}1$ baryon resonances  
is given by the PDG in Ref.~\cite{PDG1}.  
Summarizing 20 years of experimental activity  
on $S{=}{+}1$ baryons it is important to note that  
almost all searches were performed by $KN$  elastic and  
inelastic scattering at kaon momenta corresponding to   
$Z$ baryons in the mass range of 1.74${\le}M_Z{\le}$2.16~GeV.  
Furthermore, the resonance total widths reported in Ref.~\cite{PDG2}   
are very large and  
range from 70 MeV to 845~MeV. Furthermore, an expectation for  
the heavy $Z$ resonances is supported by the MIT bag  
model~\cite{MIT}, where the lightest $Z^+$ baryon has a mass of  
1.7~GeV and a smaller width.  
  
On the other hand, in the large $N_c$ limit of QCD baryons 
emerge as  soliton configurations that have to be projected on 
proper quantum numbers. In this framework the exotic  
pentaquark $Z^+$ is the lightest member of the antidecuplet of 
baryons and  arises naturally as a  rotational excitation of the 
classical  soliton (cf. Refs.~\cite{pioner}).  
For the most recent analysis of the $Z^+$ baryon properties we refer  
the reader to Refs.~\cite{Walliser,Maxim,Weigel}.  
  
The theoretical predictions for the $Z^+$ mass (and width) in the 
soliton models  vary in a wide range. Whereas the calculations in the  
Skyrme model of Ref.~\cite{Walliser} predict a $Z^+$ mass  
of 1.7~GeV, the analysis within the framework of the chiral 
quark-soliton model~\cite{Maxim}  suggests a $Z^+$ mass 
around 1.5 GeV and quite narrow width $\Gamma_Z{\leq}$15~MeV due to
specific contributions of soliton rotation (see for 
details Ref.~\cite{Maxim}).
The detailed analysis of the $Z^+$ width in the chiral  
quark soliton model indeed shows that the most favorable  
width should be about 5~MeV \cite{Rathke}. In this letter we will explore 
in particular if such a state might be detected in $NN$ scattering. 
Furthermore, we point out that  
Ref.~\cite{Weigel} predicts $M_Z$=1.58~GeV and a width 
$\Gamma_Z$=100$\pm$30~MeV. Thus the most recent  
calculations~\cite{Maxim,Weigel} suggest a low mass  
of the $Z^+$ baryon in a range that has not been investigated  
before experimentally~\cite{PDG1,PDG2}.

Here we study the possibility of  $Z^+$ ($I{=}0$, $S{=}{+}1$,  
$J^P{=}\frac{1}{2}^+$) observation in $NN$ collisions, which  
can be performed at the COoler SYchlotron (COSY) in J\"ulich.  
Furthermore, we explore the effects due to a narrow  
$Z^+$ resonance~\cite{Maxim} in the invariant mass spectra.  
  
Since the $Z^+$ resonance couples only to the $NK$ system it can be  
excited in real or virtual $K^+n$ or $K^0p$  scattering. Unfortunately,  
there are no data~\cite{PDGlast} available on the real $K^+n$ scattering  
for momenta $p_K{<}600$~GeV/c, which correspond to the mass range  
$M_Z{\le}1.58$~GeV of our interest. On the other hand,  
data exist on the $K_Lp{\to}K_Sp$ reaction~\cite{KOP} that  
can be analyzed in order to evaluate the $Z^+$ baryon properties.  
  
A virtual $Z^+$  excitation might be tested in $pp{\to}n\Sigma^+K^+$,  
$pp{\to}p\Sigma^+K^0$, $pn{\to}n{\Lambda}K^+$ and $pn{\to}p{\Lambda}K^0$  
reactions through $K$-meson exchange. Furthermore, reactions  
with a neutron in the initial state basically are performed with a  
deuteron target which might not be suitable for a measurement  
of the $Z^+$ signal in the $NK$ invariant mass; if the  
$Z^+$ is a narrow resonance, then an averaging over the  
deuteron spectral function might substantially distort the  
$Z^+$ signal. In principle this problem can be resolved by an  
additional measurement of the spectator nucleon of the deuteron  
and a full kinematical reconstruction of the final states.  
However, here we suggest to use the $K^+$ production channel, and   
also provide a motivation for the advantages of the  
$pp{\to}n\Sigma^+K^+$ reaction in searching the $Z^+$ baryon.  
  
The major uncertainty in calculations of the  
contribution from $K$-meson exchange to the $pp{\to}n\Sigma^+K^+$  
reaction is due to the poor knowledge of the $N{\Sigma}K$ coupling  
constants. The analysis of the available data only provides  
$3.5{<}g_{N{\Sigma}K}{<}6.4$~\cite{Coupl}, where the upper limit  
stems from a dispersion analysis, which might be considered as  
an almost model independent evaluation of the coupling constant from   
data. The SU(3) limit predicts $3.2{<}g_{N{\Sigma}K}{<}4.6$~\cite{SU3} 
which,  within a given uncertainty, is in reasonable agreement with the  
$N{\Sigma}K$ coupling extracted from the experimental data in 
Ref.~\cite{Coupl}.  
In the following calculations we will use $g_{N{\Sigma}K}$=3.86 as  
given by SU(3) with a mixing determined from the semileptonic hyperon  
decay~\cite{SU3}.  
  
Now, if the  $Z^+$ baryon is a narrow state, its  
contribution to the total $pp{\to}n\Sigma^+K^+$ cross section  
is proportional to the overlap between the $Z^+$ spectral function  
(taken in Breit-Wigner form) and the phase space available.  
The reaction phase space $R_3$  
increases with invariant collision energy $\sqrt{s}$  
as $R_3{\propto}\epsilon^2$, where  
$\epsilon{=}\sqrt{s}{-}m_N{-}m_\Sigma{-}m_K$. However,  
the $Z^+$ contribution saturates at energies slightly  
above $\epsilon$=$M_Z{-}m_N{-}m_K{+}3\Gamma_Z$\footnote{By  
taking three standard deviations.}, where  
$M_Z$ and $\Gamma_Z$ are the mass and the width of the  
$Z^+$ resonance, respectively. Thus for $M_Z$=1.5~GeV the threshold for   
$Z^+$ production at its pole is $\epsilon{\simeq}$66~MeV and  
the optimal ratio of the $Z^+$ contribution to the total  
production cross section -- due to other processes --   
is obtained at energies not far from threshold simply due to phase space 
arguments.  
  
Obviously, a difficulty in the observation of $Z^+$  
production in the $NN{\to}NYK$ reaction is due the 
dominance of other processes~\cite{BG} since the  
contribution from the $Z^+$ occurs only in a small part of  
the available final phase space, which lies around the  
invariant mass of the $NK$ system close to the  $Z^+$  
resonance mass. Thus it is important that the range of  
the $NK$ invariant masses dominated by $Z^+$ production is  
not affected by the intermediate $YK$ resonances and  
the $NY$ final state interaction (FSI), because they   
can  modify substantially the final observables relatively 
to the pure isotropic phase-space distributions 
as known both experimentally~\cite{COSY} and theoretically~\cite{BG}.  
  
\begin{figure}[h]  
\vspace{-10mm}\hspace{-6mm}  
\psfig{file=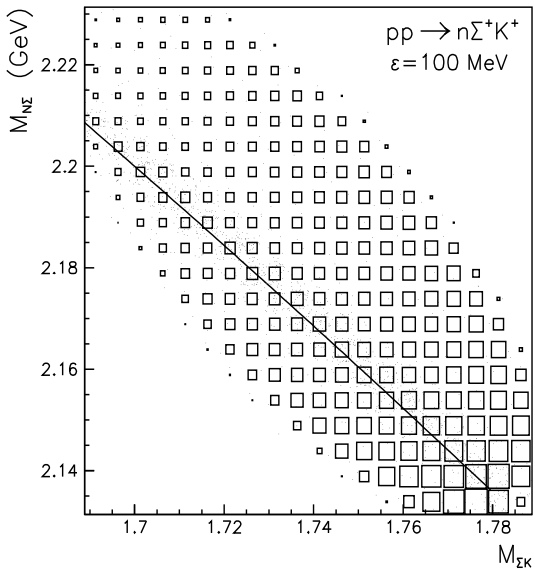,width=8.5cm,height=9cm}  
\vspace{-19mm}  
\caption{Dalitz plot for the $pp{\to}n\Sigma^+K^+$ reaction at  
an excess energy $\epsilon$=100~MeV. The size of the squares  
are proportional to the reaction cross section. The solid line shows  
the trace of the $Z^+$ pole while the distribution along the line  
is due to the $Z^+$-baryon.}  
\label{bo2}  
\vspace{-5mm}  
\end{figure}

The most dominant channel in the  
$pp{\to}n\Sigma^+K^+$ reaction is given by the  
intermediate $\Delta^{++}(1920)$ resonance. Our further  
motivations can be easily illustrated by Fig.~\ref{bo2}, that  
shows the Dalitz plot for the $pp{\to}n\Sigma^+K^+$ reaction  
at an excess energy $\epsilon$=100~MeV calculated according to  
Ref.~\cite{BG} and additionally taking into account  
the FSI~\cite{FSI,Range} as well as the $Z^+$ contribution.  
The Dalitz plot is shown as a function of the $\Sigma^+K^+$  
and $n\Sigma$ invariant masses.  
  
Furthermore, the $\Sigma^+K^+$ invariant  
mass distribution ranges from $m_\Sigma{+}m_K$ up to  
$m_\Sigma{+}m_K{+}\epsilon$, and is enhanced at large masses  
due to the $\Delta^{++}(1920)$ resonance. Notice, that large 
$\Sigma^+K^+$  
invariant masses correspond to small $n\Sigma^+$ invariant masses that  
are enhanced due to the FSI~\cite{FSI,Range}. The size  
of the squares in Fig.~\ref{bo2} is proportional to the  
production cross section.  
 
The solid line in Fig.~\ref{bo2} shows the trace of the  
$Z^+$ baryon pole with $M_Z$=1.5~GeV while the   
distribution along this line indicates the calculated  
contribution from the $Z^+{\to}nK^+$ resonance assuming a  
width $\Gamma$=5~MeV. Note that the $Z^+$ contribution  
to the $nK^+$ invariant mass spectrum at  
$M_{NK}$=$M_Z$ is obtained by integrating along the solid  
line in Fig.~\ref{bo2}, and in principle can be well  
separated from the region dominated by the $\Delta^{++}(1920)$  
resonance and the FSI by applying appropriate cuts.  
The contribution from the $Z^+$ baryon then can be detected as  
an enhancement in the Dalitz plot along the expected  
$Z^+$ resonance pole. Indeed, as illustrated by Fig.~\ref{bo2},  
such an enhancement should be detectable.  
  
\begin{figure}[h]  
\vspace{-9mm}\hspace{-6mm}  
\psfig{file=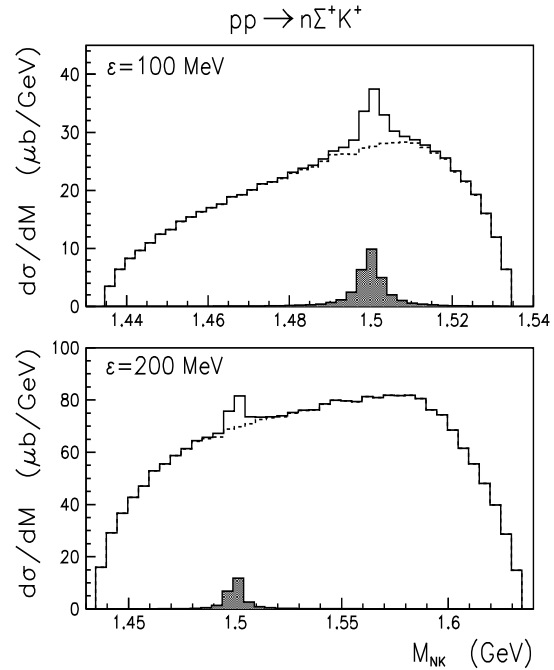,width=8.5cm,height=10cm}  
\vspace{-19mm}  
\caption{The invariant mass spectra for the $nK^+$  
system produced in the $pp{\to}n\Sigma^+K^+$ reaction at  
excess energies $\epsilon$ of 100~MeV and 200~MeV. The hatched  
histograms indicate the contribution from the $Z^+$-baryon,  
the dashed histograms show the {\it background}  
while the solid histograms display the total spectra.}  
\label{bo3}  
\vspace{-5mm}  
\end{figure}  
 
Without performing any cuts we show the invariant mass  
spectra of the $nK^+$ system produced in the $pp{\to}n\Sigma^+K^+$  
reaction at $\epsilon$=100~MeV and 200~MeV in Fig.~\ref{bo3}.  
It is important to note that the total $pp{\to}n\Sigma^+K^+$  
cross section calculated at $\epsilon$=100~MeV and 200~MeV  
amounts to 2 $\mu$b and 12.7~$\mu$b, respectively, while  
the contribution from the $Z^+$ is around 80 nb and 120~nb,  
respectively. As discussed above, the ratio of the  
$Z^+$ contribution to the total $pp{\to}n\Sigma^+K^+$ cross  
section substantially decreases with increasing  beam energy due to 
phase space.  
  
The hatched histograms in Fig.~\ref{bo3} show the $Z^+$  
contribution, the dashed histograms indicate the  
contribution from other processes, which in the following we  
discuss as {\it background} to $Z^+$ baryon production.  
The solid histograms show the total $nK^+$ invariant mass  
spectra. The contribution from the $Z^+$ baryon is well visible  
at $\epsilon$=100~MeV, while it becomes almost invisible at an 
excess energy of 200~MeV. Although the $Z^+$ contribution to the  
total $pp{\to}n\Sigma^+K^+$ cross section at $\epsilon$=100~MeV  
is very small, it can be detected in the $nK^+$ invariant mass  
spectrum in case it has a narrow width.  
It is important to note again, that lower masses of the  
$Z^+$ baryon can be detected more easily since the  
{\it background} is reduced in this case. Opposite considerations hold  
for higher $Z^+$ masses due to a larger width and background,  
respectively. 
   
\begin{figure}[h]  
\vspace{-10mm}\hspace{-6mm}  
\psfig{file=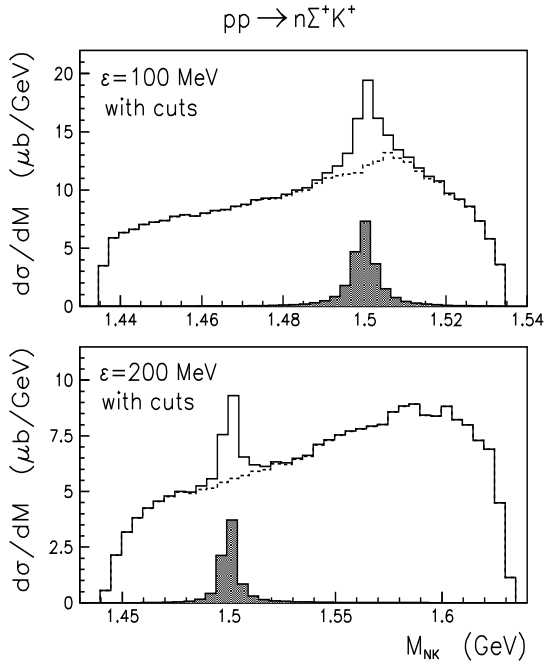,width=8.5cm,height=9.8cm}  
\vspace{-19mm}  
\caption{The same as in Fig.~\protect\ref{bo3} but with   
cuts on the $n\Sigma^+$ and $\Sigma^+K^+$ invariant masses.}  
\label{bo4}  
\vspace{-5mm}  
\end{figure}  
 
In order to improve the signal from the $Z^+$ baryon  
one can cut the regions dominated by the intermediate  
$\Delta^{++}(1920)$ resonance and the FSI. In this respect we  
show in Fig.~\ref{bo4}  
the $nK^+$ invariant mass spectra at $\epsilon$=100~MeV and  
200~MeV for the cuts, $M_{\Sigma K}{<}$1.76~MeV  
and $M_{N\Sigma}{>}$2.15~MeV. Now   
the $Z^+$ contribution becomes more pronounced even at  
an excess energy of 200~MeV.  
  
In summary, the present study indicates the possibility for an  
observation of the penta-quark $Z^+$ baryon by the $K^+n$  
invariant mass spectra, produced in  the $pp{\to}n\Sigma^+K^+$ reaction  
at excess energies around 100~MeV.  
Our estimate suggests that within an experimental statistics of  
about 200 events the $Z^+$ signal might be detected  
at $\epsilon{\simeq}$100~MeV. We note that our estimate can be  
considered conservative, because a  small $N{\Sigma}K$  
coupling constant, $g_{NK\Sigma} = 3.86$~\cite{Coupl},   
is used in the calculations. On the other hand,  
we admit that our estimate  
is based on a narrow  
$Z^+$ baryon width. If the  $Z^+$ resonance has a wider 
width~\cite{Weigel} or a higher mass, the data analysis of the 
$Z^+$ signal will  become more uncertain.  
  
\vspace{0.5cm} 
The authors like to thank W. Eyrich for his interest and encouragement 
and acknowledge the discussions on the $Z^+$ properties with   
D. Diakonov and V. Petrov.

\end{document}